\documentclass[11pt]{article}
\usepackage[pdftex]{graphicx}
\usepackage{caption}
\usepackage{subcaption}
\usepackage{dcolumn}
\usepackage{bm}
\usepackage{color}
\usepackage{amsmath}
\usepackage{stmaryrd}
\usepackage{epsfig}
\usepackage[english]{babel}
\usepackage[all]{xy}
\usepackage{geometry}
\title{Darboux transformations for supersymmetric two-boson equation }
\author{Xiao-Xing Niu$^1$, Q. P. Liu$^1$  and Lingling Xue$^2$\\[10pt]
 $^1$Department of Mathematics,\\
China University  of Mining and Technology,\\
Beijing 100083, P R China \\
$^2$Department of Mathematics,\\
Ningbo University,\\
Ningbo 315211, P R China}

\begin{document}

\maketitle
\begin{abstract}
  In this paper we construct Darboux transformations for the supersymmetric Two-boson equation. Two Darboux transformations and associated B\"acklund transformations are presented. For one of them, we also obtain the corresponding the nonlinear superposition formula.
\end{abstract}


\section{Introduction}

Broer-Kaup (BK) system or classical Boussinesq equation
\begin{equation}\label{bk}
 u_{t}=(2h+ u ^2- u_{x})_{x},\;\;\;  h_{t}=(2 u h+ h_{x})_{x},
\end{equation}
is a very important integrable system. The system was introduced by Kaup when he was extending  the newly established method of inverse scattering transform and studing an energy-dependent Sch\"ordinger spectral problem \cite{kaup0}.
Kaup further showed that the BK system may be derived from the water-wave equations if one more order of nonlinearity is included in the derivation of the Boussinesq equation \cite{kaup}.
Independently, Broer obtained the BK system as the equation approximating Boussinesq type equations \cite{broer}. As a typical integrable system, BK system has been studied extensively and a large number of results have been accumulated.
From the  viewpoint of solutions, BK system possesses various types of solutions such as the standard solitons \cite{kaup}\cite{hirota}, rational solutions \cite{nakamura}\cite{hu} and fission-fusion solutions \cite{Alonso}\cite{Satsuma}.
From the algebraic viewpoint, Kupershmidt, coined it as a dispersive water wave equation, showed that it is a tri-Hamiltonian system, namely it  has three {\em local} Hamiltonian structures which are compatible \cite{Kupershmidt}. Leo {\em et al}, within the framework of prolongation theory, worked out its B\"acklund transformation \cite{leo} and the related nonlinear superposition formula is derived by Gordoa and Conde \cite{gordoa}. The constructions of Darboux transformations for BK system were considered by Leble and Ustinov \cite{leble} and Li {\em et al} \cite{li}.
Besides, we mention that BK system also plays a role in the matrix models as explained in \cite{Aratyn}\cite{Bonora} and thus the name two-boson system.

A supersymmetric extension of BK system, named as supersymmetric two-boson system, was constructed by Brunelli and Das \cite{bd}. It reads as
\begin{equation} \label{eq1}
 \alpha_{t}=2\beta_{x}+2\alpha'\alpha_{x}-\alpha_{xx},\quad \beta_{t}=(2\alpha'\beta+\beta_{x})_{x},
\end{equation}
where $\alpha$ and  $\beta$ are fermionic or Grassmann odd variables depending on super-spatial variables $(x, \theta)$ and temporal variable $t$. $\mathcal{D}=\partial_{\theta}+\theta\partial_{x}$ is the usual super derivative and for a given function $f$,  $f'=({\mathcal D}f)$.  Brunelli and Das further demonstrated  that the supersymmetric two-boson system is a bi-Hamiltonian system and closely associated with other known integrable systems \cite{brunelli}. With Yang, one of the authors succeeded in converting the supersymmetric two-boson system into Hirota form \cite{liu}. Recently, by means of super-Bell polynomials, Fan obtained the bilinear B\"acklund transformation for it \cite{fan}.  Very recently, Xue and one of the authors examined Fan's B\"acklund transformation and worked out its related nonlinear superposition formula \cite{xue}.

The purpose of this paper is to construct Darboux transformations for the supersymmetric two-boson system.  In general, Darboux transformations have been playing vital roles in the study of integrable systems \cite{ms}. In particular,  a DT may provide a convenient tool to construct particular solutions of a (integrable) nonlinear differential equation. To the best of our knowledge, any form of Darboux transformations  for the system \eqref{eq1} has not been constructed. We also notice that the Darboux transformations and nonlinear superposition formulae are valid for the $\mathcal{N}=2,a=4$ SUSY KdV equation \cite{N2-1, N2-2}, since it shares the same hierarchy with the SUSY Two-Boson equation \cite{liu1}\cite{zhang}.

The paper is organised as follows. Next section, we will construct two Darboux transformations and the associated  B\"acklund transformations for the supersymmetric two-boson system \eqref{eq1}. We will also demonstrate the relationship bewtween our Darboux transformations and those for classical Boussinesq equation found  in \cite{li}. In Section 3, we will build the nonlinear supersymmetric formula for one of the B\"acklund transformations obtained in Section 2.

\section{Darboux-B\"acklund transformation}

It is known that  \eqref{eq1} possesses the following spectral problem
 \begin{equation}\label{eq21}
\mathcal{L}\psi = \lambda\psi,\quad
\psi_{t}= \mathcal{P}\psi,
\end{equation}
where
\begin{equation*}
\mathcal{L}=\partial_{x}-\alpha'+\mathcal{D}^{-1}\beta,\quad
\mathcal{P}=-\partial_{x}^2+2\alpha'\partial_{x}+2\beta\mathcal{D}.
\end{equation*}
The spectral problem \eqref{eq21} may be reformulated as matrix form.  To this end, we  introduce the new variable as $\Psi=(\psi,(\mathcal{D}^{-1}\beta \psi),\psi')^{\text{T}}$, and obtain
\begin{eqnarray}\label{eq3}
{\Psi}'=\mathcal{U}\Psi,\quad \mathcal{U}=\left(
    \begin{array}{ccc}
      0 & 0 & 1 \\
      \beta & 0 & 0 \\
      \lambda+\alpha' & -1 & 0 \\
    \end{array}
  \right)
\end{eqnarray}
and
\begin{eqnarray}\label{eq4}
&&\Psi_{t}=\mathcal{V}\Psi,\quad \mathcal{V}=-\left(
    \begin{array}{ccc}
     \lambda^2+\alpha_{x}'-\alpha'^2-\beta' & \alpha'-\lambda & -\beta \\
     (\lambda-\alpha')\beta'-\beta_{x}'+\beta\alpha_{x} & -\beta' & (\alpha'- \lambda)\beta+\beta_{x} \\
       (\alpha_{x}-\beta)_x-2\alpha'(\alpha_{x}-\beta)& \alpha_{x}-\beta &  \lambda^2+\alpha_{x}'-\alpha'^2-2\beta' \\
    \end{array}
  \right).
\end{eqnarray}
To construct a Darboux transformation for \eqref{eq3}\eqref{eq4}, we assume that there exists a gauge transformation 
\begin{equation}\label{eq:5}
\widetilde\Psi\equiv\mathcal{W}\Psi
\end{equation}
such that $\widetilde\Psi$ solves
\begin{equation}\label{eq:6}
\widetilde\Psi'=\mathcal{\widetilde U}\widetilde\Psi,\quad\widetilde\Psi_{t}=\mathcal{\widetilde V}\widetilde\Psi,
\end{equation}
where $\mathcal{\widetilde U}$, $\mathcal{\widetilde V}$ are the matrices ${\mathcal{U}}, {\mathcal{V}}$ but with $\alpha, \beta$ replaced by the new field variables $\widetilde\alpha,\widetilde\beta $. To be qualified as a Darboux transformation, the matrix $\mathcal{W}$ has to satisfy
\begin{equation}\label{cc}
{\mathcal W}'+\mathcal{W}^\dagger\mathcal{U}-\mathcal{\widetilde U}\mathcal{W}=0,
\quad\mathcal{W}_{t}+\mathcal{W}\mathcal{V}-\mathcal{\widetilde V}\mathcal{W}=0.
\end{equation}
It is remarked that to be consistent the gauge matrix $\mathcal{W}$ has to be structured as $\begin{pmatrix}
\text{even}& \text{even} & \text{odd}\\ \text{even}& \text{even} &\text{odd}\\ \text{odd}&\text{odd}&\text{even} \end{pmatrix}$. Also for a super matrix $A=(a_{ij})_{m\times n}$, we define $A^\dagger=(a_{ij}^\dagger)_{m\times n}$, and $a_{ij}^\dagger=(-1)^{p(a_{ij})}a_{ij}$, with $p(a_{ij})$ denoting the parity of entry $a_{ij}$.

To achieve a meaningful result, we make the simplest ansatz
 $\mathcal{W}=\lambda\mathcal{M}+\mathcal{N},\; \mathcal{M}=(m_{ij})_{3\times 3},\; \mathcal{N}=(n_{ij})_{3\times 3}$.
It follows from the first equation of \eqref{cc}  that the matrices $\mathcal{M}$ and $ \mathcal{N}$ should take the following forms:
\[
\mathcal{M}=\begin{pmatrix}m_{11}&0&0\\ 0&m_{22}&0\\ m_{31}&0&m_{11}\\ \end{pmatrix},
\; \;
\mathcal{N}=\left(\begin{array}{ccc}n_{11}& m_{22}-m_{11} &m_{11}'-m_{31} \\
n_{21}&n_{22}&m_{22}\beta-m_{11}\widetilde\beta\\\
n_{31}&-m_{31}&n_{33}\\
\end{array}\right),
\]
where the entries are determined by the following equations
\begin{align}
m_{22}'&=0,\label{1}\\
n_{22}'+m_{22}(\beta-\widetilde\beta)&=0,\label{7}\\
n_{11}-n_{33}-m_{31}'+m_{11,x}&=0,\label{3}\\
n_{33}-n_{11}+m_{31}'+m_{11}\left(\alpha'-\widetilde\alpha'\right)&=0,\label{2}\\
n_{22}-n_{33}-m_{31}'-(m_{22}-m_{11})\widetilde\alpha'&=0,\label{6}\\
n_{21}+(m_{22}\beta-m_{11}\widetilde\beta)'-\widetilde\beta(m_{11}'-m_{31})&=0,\label{8}\\
n_{31}-n_{11}'-(m_{22}-m_{11})\beta+(m_{11}'-m_{31})\alpha'&=0,\label{4}\\
n_{33}'-n_{31}-(m_{11}'-m_{31})\widetilde\alpha'+m_{22}\beta-m_{11}\widetilde\beta&=0,\label{9}\\
n_{31}'+n_{21}+n_{33}\alpha'-n_{11}\widetilde\alpha'+m_{31}\beta&=0,\label{11}\\
n_{21}'+n_{22}\beta-n_{11}\widetilde\beta-(m_{22}\beta-m_{11}\widetilde\beta)\alpha'&=0. \label{10}
\end{align}
It follows from \eqref{1}  that $m_{22}$ is a constant. Without loss of generality, one may consider $m_{22}=0$ or $m_{22}=1$ . Also, \eqref{2} \eqref{3} yield
\begin{equation}\label{12}
m_{11,x}+m_{11}(\alpha'-\widetilde\alpha')=0.
\end{equation}

Now, we consider the following cases:\\

{\bf Case I:}
We assume $m_{11}=0,\; m_{22}=1$ and replace $n_{11}$  by $-{r}$, for simplicity.
Then the equations  \eqref{2}-\eqref{4} lead to
\begin{align*}
&n_{33}=-{r}-m_{31}',\qquad n_{21}=-\beta'-\widetilde\beta m_{31},\\
&n_{22}=-{r}+\widetilde\alpha',\;\;\qquad\; n_{31}=-{r}'+\beta+m_{31}\alpha'.
\end{align*}
Above equations together \eqref{9} produce
\begin{equation*}
m_{31}(\widetilde\alpha'-\alpha')-m_{31,x}=0,
\end{equation*}
which holds if we  choose $m_{31}=0$. Now \eqref{11} gives
\[r_x=r(\widetilde\alpha-\alpha)'.\]
Next supposing $r\neq 0$, integrating above equation once, and setting the integral constant to be zero, without loss of generality,
one has
\begin{equation}\label{101}
\widetilde\alpha=\alpha+(\ln {r})'.
\end{equation}
Then \eqref{7} supplies us with
\begin{equation*}
\widetilde\beta=\beta+\alpha_{x}-{r}'+(\ln {r})_{x}'.
\end{equation*}
Finally, \eqref{10} gives the equation
\begin{equation*}
\alpha_x-(\beta/r)_x-r'+(\ln r)'_x=0.
\end{equation*}
which may be integrated once and yields
\begin{equation}\label{bt1}
r_{x}={r}^2-\lambda_1{r}-{r}\alpha'+r({\beta /{r}})'.
\end{equation}
where $\lambda_1$ is an integral constant.
Above equation constitutes a (spatial part) BT for the  SUSY Two-Boson equation. The associated temporal part of BT may be obtained as follows
\begin{equation}\label{bt1-t}
r_{t}=(r_{x}+2{r}\alpha'-{r}^2)_{x}.
\end{equation}
To sum up, Darboux transformation for the field variables reads as
\begin{align}\label{DT1}
\widetilde\alpha =\alpha+(\ln {r})', \quad \widetilde\beta& =\beta+\left({\beta}/{{r}}   \right)_x, 
\end{align}
and  Darboux transformation   for the eigenfunctions  reads as
\begin{align}\label{103t}
\widetilde\Psi=\mathcal{W}_{I}\Psi,\quad \mathcal{W}_{I}=\left(
    \begin{array}{ccc}
      -{r} & 1 & 0 \\
     -\beta' & \lambda-\lambda_1+\left({\beta}/{{r}}\right)' & \beta \\
      \beta-{r}' & 0 & -{r} \\
    \end{array}
  \right),
\end{align}
together with \eqref{bt1}-\eqref{bt1-t}, which define the B\"acklund transformations for the field variables.

\noindent
{\bf Remark 1}:
 On the level of eigenfunctions, the Darboux transformation in scaler form is easily found to be:
\[\widetilde{\psi}={\cal D}^{-1}(\beta\psi)-{r}\psi.\]

{\bf Case II:}
As we know that the inverse of a Darboux matrix is also a Darboux matrix, thus we may construct the second Darboux tranformation from the case I. In fact,
we consider the following  replacement:
\begin{equation*}
(\lambda_1,\; \alpha,\;\beta,\;\widetilde\alpha,\;\widetilde\beta,\;r)\rightarrow(\lambda_2,\;\widetilde\alpha,\;\widetilde\beta,\; \alpha,\; \beta,\;s),
\end{equation*}
then it follows from \eqref{bt1} and  \eqref{DT1} that
\begin{align}\label{bt2}
s_{x}=&{s}^2-\lambda_2{s}-{s}\widetilde\alpha'+{s}(\widetilde\beta'/{s})',\\
\label{DT21}
\widetilde\alpha=&\alpha-(\ln {s})' ,\\
\label{DT22}
\widetilde\beta=&\beta-\left(\widetilde\beta/{s}   \right)_x. 
\end{align}
Substituting \eqref{DT21} into \eqref{bt2}, one has
\begin{equation}\label{btx0}
(\widetilde\beta/s)'=\alpha'+\lambda_2-s.
\end{equation}
From above equation together with \eqref{DT22}, one obtains
\begin{align}\label{DT23}
 \widetilde\beta=\beta-\alpha_x+{s}' 
\end{align}
and
\begin{equation}\label{btx}
s_{x}={s}(\lambda_{2}+\alpha')-{s}^2+{s}\Big({s}^{-1}(\alpha_{x}-\beta)\Big)'.
\end{equation}
Thus \eqref{DT21} and \eqref{DT23} are the Darboux transformation for the field variables, \eqref{btx} serves as the spatial part of B\"acklund transformation.
By direct calculation, we find the temporal part of BT as follows
\begin{equation*}
s_{t}=-s_{xx}-({s^2}- 2{s}\alpha')_x.
\end{equation*}

Now we construct the corresponding Darboux matrix. Considering the matrix $(\lambda-\lambda_1)W_{I}^{-1}$, replacing $\lambda_1,\;\beta,\;r$ by
$\lambda_2,\;\widetilde\beta,\;s$,
together with \eqref{btx0} and \eqref{DT23}, one obtains the second Darboux transformation for the wave function
\begin{align*}
\widetilde\Psi={\mathcal{W}}_{II}\Psi,\quad \mathcal{W}_{II}=-\frac{1}{s}\left(
    \begin{array}{ccc}
      \lambda -{s}+\alpha' & -1 & (\zeta-{s}')/{s}\\
    -{{s}^2} +{s}(\lambda_{2}+\alpha') &-{s} & \zeta-{s}' \\
      \zeta-(\lambda +\alpha')\zeta/{s} & \zeta/{s} & \lambda -\lambda_{2}-{s}'\zeta /{s^2}\\
    \end{array}
  \right),
\end{align*}
where
\begin{align*}
\alpha_{x}-\beta=\zeta.
\end{align*}

\noindent
{\bf Remark 1}: Up to a simple change of variables, the BT in the Case II coincides with the one appeared in the framework of super-Bell polynomials (see \cite{fan} or  \cite{xue}).\\
{\bf Remark 2}:
One may rewrite the transformation as a scaler form:
\[\widetilde{\psi}=\psi-{s}^{-1}\psi_x+{s}^{-2}({s}'-\alpha_{x}+\beta)\psi'\]

So far, we have constructed two different Darboux transformations for \eqref{eq3} and the related B\"acklund transformations for the supersymmetric two-boson system. As usual, it is interesting to work out the Darboux matrices in more explicit forms, namely, to represent the entries of Darboux matrices in terms of the solutions of the linear spectral system or its adjoint.

In {\bf Case I},  a direct calculation shows that the Berezian or super-determinant of ${\cal W}_{I}$ reads as
\[
Ber(\mathcal{W}_{I})=\lambda-\lambda_{1},
\]
thus take a particular solution $\Psi_1=(f_1,g_1,\sigma_1)^{\text{T}}$ of the spectral problem \eqref{eq3} at $\lambda=\lambda_1$. Here, $f_1$ and $g_1$ are bosonic, $\sigma_1$ is fermionic. Imposing ${\cal W}_{I}\Psi_{1}|_{\lambda=\lambda_{1}}=0$, we have
\begin{equation*}
{r}={g_1}/{f_1}.
\end{equation*}
In this way, the deserved explicit form of the Darboux matrix $\mathcal{W}_{I}$ is found.

For the {\bf Case II}, we need the adjoint linear problem of \eqref{eq3},  which is  given by
\begin{equation}\label{eq101}
-(\chi^\dag)'=\chi \mathcal{U}.
\end{equation}
Now since each  Darboux transformation for the linear spectral problem induces a   Darboux transformation
for the adjoint linear spectral problem, we have
\[
\widetilde\chi=\chi \mathcal{T}
\]
with
\begin{align*}
\mathcal{T}&=(\lambda-\lambda_{2})(\mathcal{W}_{II}^{-1})^\dag\\
&=\left(
    \begin{array}{ccc}
     -{s} & 1 & 0 \\
      {s}^2-{s}(\lambda_2+\alpha')+{s}'{s}^{-1}\zeta &\lambda-{s}+\alpha'&\zeta-{s}'\\
      \zeta & 0 & -{s} \\
    \end{array}\right).
\end{align*}
Note that
\[
Ber(\mathcal{T})=\lambda-\lambda_{2}.
\]
Thus, we take a particular solution $\chi_{2}=(g_{2},f_{2},\sigma_{2})$ of \eqref{eq101} at $\lambda=\lambda_{2}$, where $f_{2}$ and $g_{2}$ are bosonic, $\sigma_{2}$ is fermionic. By imposing $\chi_{2}{\cal T}|_{\lambda=\lambda_{2}}=0$, we obtain
\begin{equation*}
{s}=\lambda_{2}+\alpha'+{g_{2}}/{{f_{2}}}.
\end{equation*}
With this result, the explicit form of Darboux matrix ${\mathcal{W}}_{II}$ is achieved.

As a final part of the section, we relate our Darboux transformations to those for the classical Boussinesq equation obtained in  \cite{li}.
Then letting $\alpha=\xi+\theta  u $ and $\beta=\eta+\theta h $, from \eqref{eq3} we find we easily obtain the bosonic limit
\begin{eqnarray}\label{das1}
\varphi_{x}=U_{1}\varphi,\;\;U_{1}=\left(
    \begin{array}{cc}
     \lambda+ u  & -1  \\
      h & 0 \\
    \end{array}
  \right).
\end{eqnarray}
From \eqref{eq4}, we have
\begin{eqnarray}\label{das2}
\varphi_{t}=V_{1}\varphi,\;\;V_{1}=-\left(
    \begin{array}{cc}
     \lambda^2- u ^2-h+ u_{x} & u-\lambda   \\
      \lambda h- u h- h_{x} & -h \\
    \end{array}
  \right).
  \end{eqnarray}
The compatibility of \eqref{das1} and \eqref{das2} gives \eqref{bk}.
 By taking the gauge transformation
\begin{eqnarray*}
\varphi=G\phi,\quad G=g\left(
    \begin{array}{cc}
     1 & 0  \\
     0 &-1 \\
    \end{array}
  \right),
  \end{eqnarray*}
  where $g$ satisfies the following equation
\[ g_x=\frac{1}{2}(\lambda+u )g,
\]
the spectral problem \eqref{das1}  
becomes
\[
\phi_x=M\phi,\quad M=\left(
    \begin{array}{cc}
     (u-\lambda)/2 & 1  \\
     -h &-(u+\lambda)/2 \\
    \end{array}
  \right)
\]
which is essentially the one considered in \cite{li} (cf. (4)  of \cite{li}).
The bosonic limit of the Darboux matrix ${\mathcal{W}_{I}}$ reads as
\[
W_1=\begin{pmatrix}
-a&1\\ -h& \lambda-\lambda_1+\frac{h}{a}
\end{pmatrix} 
\]
with 
\[\widetilde{u}=u+(\ln a)_x,\quad \widetilde{h}=h+(h/a)_x,\]
where $a$ is defined by
\[
a_{x}={a}^2-\lambda_1{a}-{a}{u}+{h}.
\]
Then, direct calculation shows that the  Darboux transformation
\[
\widetilde\phi=T\phi,\quad
T=\widetilde{G}^{-1} W_1 G=d\begin{pmatrix}
-a&-1\\ h& \lambda-\lambda_1+\frac{h}{a}
\end{pmatrix}, \quad a d^2=-1\;\; 
\]
which is depicted by
\begin{displaymath}
    \xymatrix{ \varphi \ar[d]_{W_1}& \phi\ar[l]_{G} \ar@{.>}_{T}[d] \\
               \widetilde{\varphi} \ar[r]_{\widetilde{G}^{-1}} & \widetilde\phi }
\end{displaymath}
 is exactly the Darboux transformation described
by the Proposition 2 of \cite{li}.


Similarly, it can be shown that the bosonic limit of  our Darboux transformation presented in Case III relates to the Darboux transformation presented by Proposition 1 of \cite{li}.
%
%
%
%
\section{Nonlinear Superposition Formula}
For a given a B\"acklund transformation, it is interesting to find the corresponding nonlinear superposition formula. For the {\bf Case II} of the last section, such formula has been worked out already in \cite{xue}. In this section, we consider the nonlinear superposition formula for the B\"acklund transformation given in the {\bf Case I}. Thus, we start with a solution $(\alpha, \beta)$ of \eqref{eq1} and take two solutions $\Psi_1=(f_1,g_1,\sigma_1)^{\text{T}}$ and $\Psi_2=(f_2,g_2,\sigma_2)^{\text{T}}$ corresponding to arbitrary constants $\lambda_1$ and $\lambda_2$. Then we consider two Darboux transformations
\begin{eqnarray}
  \Psi_{[k]} &=& \mathcal{W}_k\Psi,\quad \mathcal{W}_k\equiv {\mathcal{W}}_{II}|_{\lambda=\lambda_k}=\left(
    \begin{array}{ccc}
    - r_{k} & 1 &0  \\
      -\beta' & \lambda-r_{k}+\alpha'+\frac{r_{k,x}}{r_{k}} & \beta \\
   \beta- r_{k}'& 0 &-r_{k}\\
    \end{array}
  \right),
\end{eqnarray}
and
\[
\alpha_{[k]}=\alpha+(\ln r_{k})', \;\; \beta_{[k]}=\beta+(\frac{\beta}{r_{k}})_{x},\;\; r_k=\frac{g_k}{f_k}.
\]
Then from the compatibility of the two equations, or the Bianchi identity

\begin{displaymath}
     \xymatrix{
                &&(\alpha_{[1]},\beta_{[1]}; \Psi_{[1]})\ar[drr]^{\lambda_2}&& \\
 (\alpha,\beta; \Psi) \ar[urr]^{\lambda_1} \ar[drr]_{\lambda_2} && && (\alpha_{[12]},\beta_{[12]}; \Psi_{[12]})=  (\alpha_{[21]},\beta_{[21]}; \Psi_{[21]})  \\
                             && (\alpha_{[2]},\beta_{[2]}; \Psi_{[2]})\ar[urr]_{\lambda_1}&&
}\end{displaymath}
we have
\[
{\mathcal W}_{2{[1]}}{\mathcal W}_1={\mathcal W}_{1{[1]}}{\mathcal W}_2,
\]
where
 \begin{align*}
 \mathcal{W}_{1[1]}&={}\mathcal{W}_{1}|_{\alpha=\alpha_{[2]},\beta=\beta_{[2]},r_{1}=r_{21}},\;\; \quad\mathcal{W}_{2[1]}=\mathcal{W}_{2}|_{\alpha=\alpha_{[1]},\beta=\beta_{[1]}, r_{2}=r_{12}},\\
  \alpha_{[21]}&={}\alpha_{[2]}+(\ln r_{21})',\;\;\;\;\;\;\;\;\alpha_{[12]}=\alpha_{[1]}+(\ln r_{12})',\\
  \beta_{[21]}&={}\beta_{[2]}+(\frac{\beta_{[2]}}{r_{21}})_{x},\;\;\;\;\;\;\;\;\beta_{[12]}=\beta_{[1]}+(\frac{\beta_{[1]}}{r_{12}})_{x}.
 \end{align*}
From~$\alpha_{[12]}=\alpha_{[21]}, \beta_{[12]}=\beta_{[21]},\Psi_{[12]}=\Psi_{[21]}$, we can get
\begin{equation*}
\alpha_{[12]}=\alpha+(\ln r_{1}r_{12})',\quad\quad\quad\beta_{[12]}=\beta+\left(\frac{\beta}{r_{1}}+\frac{\beta+(\frac{\beta}{r_{1}})_{x}}{r_{12}}\right)_x,
\end{equation*}
where
\begin{equation*}
r_{12}=\frac{(\lambda_{2}-\lambda_{1})r_{2}}{r_2-r_1}+\frac{\beta'}{r_{1}}+\frac{\beta r_{2}}{r_2-r_1}\left(\frac{r_1'}{r_{1}^2}-\frac{r_{2}'}{r_{2}^2}\right).
\end{equation*}

As a final remark, we notice that our results an Darboux transformations and nonlinear superposition formulae are valid for the $\mathcal{N}=2,a=4$ SUSY KdV equation \cite{N2-1, N2-2} , since it shares the same hierarchy with the SUSY Two-Boson equation \cite{liu1}\cite{zhang}.

\section*{Acknowledgments}
This paper is supported by the National Natural Science Foundation of China (grant numbers: 11271366, 11331008 and 11401572) and the Fundamental Research Funds for Central Universities.

\end{document}